\documentstyle[aas2pp4,amstex,epsfig,rotating,float,11pt]{article}

\def\refindent{\par\noindent\hangindent=3pc\hangafter=1 }
\def\aa#1#2#3{\refindent#1, A\&A, #2, #3}
\def\aasup#1#2#3{\refindent#1, A\&AS, #2, #3}

\def\apj#1#2#3{\refindent#1, {\it ApJ}, {\bf#2}, #3.}
\def\apjlett#1#2#3{\refindent#1, {\it ApJ (Letters)}, {\bf #2}, #3.}
\def\apjsup#1#2#3{\refindent#1, ApJS, #2, #3}

\def\mnras#1#2#3{\refindent#1, {\it MNRAS}, {\bf#2}, #3.}
\def\nature#1#2#3{\refindent#1, {\it Nature}, {\bf #2}, #3.}

\def\sovlett#1#2#3{\refindent#1, Soviet Astr.\ Lett., #2, #3}

\def\>{$>$}
\def\<{$<$}

\def\simlt{\lower.5ex\hbox{$\; \buildrel < \over \sim \;$}}
\def\simgt{\lower.5ex\hbox{$\; \buildrel > \over \sim \;$}}
\def\sqr#1#2{{\vcenter{\hrule height.#2pt
      \hbox{\vrule width.#2pt height#1pt \kern#1pt
         \vrule width.#2pt}
      \hrule height.#2pt}}}

\begin{document}

\centerline{Astrophysical Journal (Letters), in press}
\bigskip
\title{On The Nature Of The EGRET Source\\
    At The Galactic Center}

\author{Sera Markoff\altaffilmark{1}$^*$, Fulvio Melia\altaffilmark{2}$^{*\dag}$
and Ina Sarcevic$^*$}
\affil{$^*$Physics Department, The University of Arizona, Tucson, AZ 85721}
\affil{$^{\dag}$Steward Observatory, The University of Arizona, Tucson, AZ 85721}





\altaffiltext{1}{NSF Graduate Fellow.}
\altaffiltext{2}{Presidential Young Investigator.}


\begin{abstract}
The recent detection of a $\gamma$-ray flux from the direction of the
Galactic center by EGRET on the Compton GRO raises the question of
whether this is a point source (possibly coincident with the massive
black hole candidate Sgr A*) or a diffuse emitter.  Using the latest
experimental particle physics data and theoretical models, 
we examine in detail the $\gamma$-ray spectrum
produced by synchrotron, inverse Compton scattering and mesonic decay
resulting from the interaction of relativistic protons with hydrogen
accreting onto a point-like object.  Such a distribution of high-energy
baryons may be expected to form within an accretion shock as the
inflowing gas becomes supersonic.  This scenario is motivated by
hydrodynamic studies of Bondi-Hoyle accretion onto Sgr A*, which
indicate that many of its radiative characteristics may ultimately be
associated with energy liberated as this plasma descends down into the
deep potential well.  Earlier attempts at analyzing this process
concluded that the EGRET data are inconsistent with a massive
point-like object.  Here, we demonstrate that a more careful treatment
of the physics of $p$-$p$ scattering suggests that a $\sim 10^6\;M_\odot$
black hole may be contributing to this high-energy emission.

\end{abstract}


\keywords{acceleration of particles---black hole physics---Galaxy: 
center---galaxies: nuclei---gamma rays: theory---radiation mechanisms:
non-thermal}


%

\section{Introduction}
X-ray and $\gamma$-ray emission have been detected from the
Galactic center (\cite{wat81}; \cite{skin87}; \cite{pred94};
\cite{chu94}).  The implications for the radio point source Sgr A* are
rather interesting, since the X-ray luminosity is not as large as what
is expected based on X-ray observations of smaller black hole
candidates.  Recently, EGRET on board the Compton GRO has
identified a central ($< 1^o$) $\sim 30$ MeV - $10$ GeV continuum
source with luminosity $\approx 5\times 10^{36}$ ergs s$^{-1}$
(\cite{mat96}; \cite{mae96}; \cite{mer96}).  This EGRET $\gamma$-ray
source (2EGJ1746-2852), appears to be positioned at $l\approx 0.2^o$,
but the exact center of the Galaxy, or even a negative longitude, 
cannot be ruled out completely.  Its spectrum can be fit by a
hard power-law of spectral index $\alpha=-1.74\pm 0.09$
($S=S_0\,E^\alpha$), with a cutoff between $4-10$ GeV.  At lower
energies, the COMPTEL data provide useful upper limits (\cite{str96}).

These $\gamma$-rays may originate either (1) close to the massive
black hole, possibly from relativistic particles
accelerated by a shock in the accreting plasma (\cite{mo94}), or
(2) in more extended features where relativistic particles are known
to be present (\cite{po97}).  
In the former (see also \cite{mah97}, who considered a thermal distribution
of hot protons), the $\gamma$-rays may result from the decay of 
$\pi$'s produced via $p$-$p$ collisions of ambient protons with the
shock-accelerated relativistic protons.  This study concluded that the
presence of a $M_h\sim 10^6\;M_\odot$ black hole was inconsistent with
the EGRET data.  However, the earlier calculations suffered from
over-simplification and an incomplete treatment of the physics of
$p$-$p$ scattering.  For example, although the multiplicity of $\pi$
production (i.e., the number of $\pi$'s produced per collision) is a
strong function of energy, it was approximated with a constant value
of 3 when in fact it can change by orders of magnitude.  
In addition, ignoring the role of cascading protons is not a
valid approximation when the energy carried away by the leading
protons can be over $90\%$ of the incoming proton energy.  Here we
examine the hypothesis of a black hole origin for the
$\gamma$-rays employing the most current data for the energy-dependent
cross-sections, inelasticity, and $\pi$ multiplicity, together with a
self-consistent treatment of the particle cascade.  

\section{The Physical Picture}
Sgr A* is a nonthermal radio source at the center of
the Galaxy (e.g., \cite{men97}).  The large mass ($\sim 1.8\times10^6
M_\odot$; \cite{ha96}; \cite{eg97}) enclosed within $\sim 0.1$ pc
indicates that this object may be a massive black hole.  The nearby
IRS 16 cluster of hot giant stars produces a Galactic
center wind with velocity $v_{gw}\approx500-700$ km
s$^{-1}$ and mass-loss rate $\dot{M}_{gw}\approx3-4\times10^{-3}
\;M_\odot$ yr$^{-1}$.  A portion of this wind is captured by 
Sgr A* and accretes inward.  The accretion rate
($\simlt 10^{22}$ g s$^{-1}$) resulting from this Bondi-Hoyle process
is well below the Eddington value, and so the gas attains a free-fall
velocity (\cite{mel94}; \cite{ruf94}), which eventually becomes
supersonic, and a shock forms at $r_{sh}\sim 40-120\, r_g$
(\cite{ba89}), where $r_g\equiv{2GM_{h}}/{c^2}$ is the Schwarzschild
radius.  A fraction of the particles may be accelerated to very 
high energy by the shock.
However, the greater synchrotron and inverse Compton efficiency of the
$e^-$'s compared to that of the $p$'s limits the maximum attainable
Lorentz factor of the former by several orders of magnitude compared
to the latter, and so the (accelerated) $e^-$ contribution to
the radiation field is negligible.  The relativistic $p$'s are
injected through the shock region with a rate
$\dot{\rho}_p(E_p)=\rho_oE_p^{-x}$ cm$^{-3}$ s$^{-1}$ GeV$^{-1}$.  In
steady state, this leads to a power-law distribution with
index $z\sim 2.0-2.4$ (\cite{je91}).  In our case, $z$ is determined
in part by the $p$ cooling processes and the particle cascade, and as
we shall see, $z\simgt x$.  The normalization $\rho_o$ is related to
the efficiency $\eta$ of the shock by
\begin{equation}
\int \rho_o E_p^{-x} E_p\;
dE_p=\eta L_{grav}\equiv\frac{\eta GM_h\dot{M}}{r_{sh}}\;.
\end{equation}

These relativistic particles interact with the ambient particles and
the magnetic field $B$, producing photons via synchrotron, inverse
Compton scatterings and the decay of mesons created during $p$-$p$
collisions.  Because the injected $p$'s can be ultrarelativistic,
leading order nucleons produced in the scattering events also
contribute to the spectrum via multiple collisions in an ensuing
cascade.  The main products in these collisions are $\pi$'s, which
then decay either to photons ($\pi^0\rightarrow\gamma\gamma$) or
leptons ($\pi^\pm\rightarrow \mu^\pm\nu_\mu$, with $\mu^\pm\rightarrow
e^\pm\nu_e \nu_\mu $).

Following Melia (1994), one can see that on average the ambient $p$ 
number density is 
\begin{equation}\label{pdis}
n_p=\frac{\dot{N}}{4\pi
cr_g^2} \left(\frac{r_g}{r}\right)^{3/2}\;,
\end{equation} 
where $\dot{N}\approx \dot{M}/m_p$ is the $p$ number accretion rate
in terms of the mass $m_p$.
If in addition the magnetic field $B$ is in approximate equipartition with
the kinetic energy density, then
\begin{equation}\label{bfield}
B^2=\left(\frac{\dot{M}c}{r_g^2}\right)\left(\frac{r_g}{r}\right)^{5/2}\;.
\end{equation}
The frequency $\nu_{max}$ at which the gas becomes transparent 
lies in the Rayleigh-Jeans portion of the spectrum.
This radiation is assumed to be in
thermal equilibrium with the $e^-$'s.

\section{Properties of the Particle Cascade}
The relativistic $p$'s undergo a series of interactions including 
$p N\rightarrow p N\, M_\pi\, M_{N\bar N}$,
where $N$ is either a $p$ or a neutron $n$, $M_\pi$ represents the 
energy-dependent multiplicity of $\pi$'s, and $M_{N\bar N}$ is the multiplicity of 
nucleon/anti-nucleon pairs (both increasing functions of energy).  
Since $M_{N\bar N}/M_\pi< 10^{-3}$ at low energy and even smaller
at higher energies (\cite{cli88}), we here ignore
the anti-nucleon production events. The charge exchange interaction
($p\rightarrow n$) occurs roughly $25\%$ of the time (e.g., \cite{be90}).  
The other possible interactions are 
$p\gamma\rightarrow p\pi^0 \gamma$, $p\gamma\rightarrow n \pi^+ \gamma$,
$p \gamma\rightarrow e^+e^-p$ and $p e \rightarrow e N M_\pi$.

The high-energy cutoff for the injected $p$ distribution is set by
determining the Lorentz factor $\gamma_{p,max}$ above which the combined energy loss
rate due to synchrotron emission, inverse Compton scattering and
hadronic collisions exceeds the rate of energy gain due to shock
acceleration.  This transition energy depends 
on the functional form of the inelasticity and the fraction of 
power transferred to the $\pi$'s during the $p$-$p$ collisions.  

Using the $M_\pi$ measured at several center-of-momentum (CM) energies
(\cite{alp75}; \cite{abe88}; \cite{alb90}), one can
determine the $\pi$ injection rate from the $p$ distribution and the
physical characteristics of the ambient medium.  The
particle cascade continues with the emission of $\gamma$-rays and
leptonic decays.  The $e^-$'s and $e^+$'s produced in this fashion
constitute an energetic population and one must assess
their contribution to the spectrum via synchrotron emission
and inverse Compton scattering.

For the conditions in Sgr A*, the $p$-$p$ collisions dominate over all
other $\pi$ production modes.  The relevant energy ($E_p$) range is
bounded below by the $\pi$ production threshold and above
by $\gamma_{p,max}$.  Using logarithmic bins, and assuming time independence,
we first calculate the steady state $p$ distribution $\rho_p(E_p)$
using the diffusion loss equation
\begin{multline}\label{pdens}
\rho_p(E_p)=\left[.892\int
R_{pp}(E'_p)\,\Delta\,dE'_p+\dot{\rho}_p(E_p)\right.\\
-\left.\kappa_p\,E_p^2\,\frac{\partial\rho_p}
{\partial E_p}\right]\left.\right/\left[n_p\sigma_{pp}(E_p)c+2E_p\,\kappa_p\right]\;,
\end{multline}
where $\Delta\equiv\left[\delta(E'_p-\bar{E}_{p,1})+\delta(E'_p-\bar{E}_{p,2})\right]$ and
$R_{pp}(E_p)=n_p\sigma_{pp}(E_p)c\rho_p(E_p)\,\,\text{cm}^{-3}\text{s}^{-1}\text{GeV}^{-1}$
is the rate of $p$-$p$ collisions at energy $E_p$, and
$\kappa_p=(4\sigma_Tm_e^2/3c^3m_p^4)(U_B+U_{\gamma})$ is the sum of
constants appearing in the power $P_{sync}+P_{Compton}\approx \kappa_p\, E_p^2$.  The
first term represents the influx of $p$'s having
energy $E_p$ as secondaries in $p$-$p$ collisions
between relativistic $p$'s with energies $\bar{E}_{p,1}$ and
$\bar{E}_{p,2}$ and the ambient $p$'s (which are effectively at
rest).  The relationship between $E_p$ and $\bar{E}_{p,1}$ and
$\bar{E}_{p,2}$ is unique, as expressed by the delta-functions, and is
determined by special relativity and the inelasticity $K_{pp}$.  We
use the assumption that on average the two leading $p$'s created
travel either parallel or opposite to the boost to simplify our
calculations.  In the $p$-$p$ CM frame, we use $K_{pp}=1.35\, s^{-0.12}$
for $\sqrt{s} \ge 62\; \text{GeV}$, and $K_{pp}=0.5$ for $\sqrt{s} \le
62\; \text{GeV}$, where the higher energy slope is from Alner et al. (1986),
normalized to match the approximately constant low energy value
(\cite{fo84}).

The cross section $\sigma_{pp}$ is taken as a function of energy from
the most current published data (\cite{ba96}).  Unfortunately, the
highest energy achieved in modern colliders is orders of magnitude
below the values attained in our system.  However, the data for
$\sqrt{s}\simgt 100$ GeV have a log-linear form which makes it possible to
extrapolate up to much higher $E_p$.  For the entire range, this is
within the Froissart upper bound, which states that at extremely high
energy, $\sigma_{pp\;\infty}\propto (\ln s)^2$.

The steady state relativistic $p$ distribution resulting from this
procedure can then be used to calculate the synchrotron and inverse
Compton scattering spectra, following Rybicki \& Lightman (1979), and
the rate $R_{pp}$ of $p$-$p$ collisions.  For each of these
collisions, a multiplicity $M_\pi$ of $\pi$'s is produced, with a
ratio of charged to neutral particles of roughly 2:1 (exactly when
there are two leading $p$'s, otherwise there will be a surplus $\pi^+$
to conserve charge).  These have a distribution in transverse (to the
beam in experiments, in our case to the direction of the boost back to
the lab frame) momentum ${dN_\pi/dp_\perp}$, which is measured as a
function of $\sqrt{s}$ at collider experiments.  In order to find the
energy of the $\pi$'s in the CM frame, we also need the parallel
component of the momentum, $p_\parallel$, which we extract from the
$\pi$ distribution as a function of the rapidity, $y$.  In the CM
frame, $y=(1/2)\ln[(E^*_\pi + p_\parallel)/(E^*_\pi-p_\parallel)]$,
and $y\approx-\ln[\tan(\theta/2)]$ for relativistic energies, where
$\cos\theta=p_\parallel/\mid p\mid$.  At lower energy ($\sqrt
s\simlt200$ GeV), ${dN_\pi/dy}$ is Gaussian in shape, the top of which
widens gradually into a plateau with increasing energy.  The width and
the height of this plateau can be fit to functional forms in
$\sqrt{s}$.  Details of this process are discussed in an upcoming
paper.  After binning in $p_\perp$, the end product is a distribution
${dN_\pi/dp_\perp dy}$ of $\pi$'s in the CM frame, for each $p$-$p$
collision at any particular CM energy.  Given $y$ and $p_\parallel$,
and the fact that $E^*_\pi=(m_\pi^2+ p_\parallel^2+p_\perp^2)^{1/2}$,
the distribution ($\text{cm}^{-3} \text{s}^{-1}$ per $p$-$p$ collision
at $E_p$) of $\pi$'s at energy $E^*_\pi$ follows from a convolution of
$(dN_\pi/dp_\perp dy) dp_\perp dy$ with $R_{pp}$.  

Each of the photons produced in the $\pi^0$ decay acquires a rest
frame energy $\epsilon_\gamma'=(1/2)m_\pi c^2$ and is emitted with
equal probability in any direction. The photon number density
in the observer's (or lab) frame must therefore be
\begin{equation}\label{flat}
{\rho}_\gamma(\epsilon_\gamma)\;d\epsilon_\gamma=
\frac{2\,d\epsilon_\gamma}{\beta_\pi\gamma_\pi m_\pi c^2}\;,
\end{equation}
where $\gamma_\pi$ and $\beta_\pi$ are the $\pi$ Lorentz factor
and dimensionless velocity, respectively, in this frame.
Finally, the photon emissivity is given by the multiple integral expression
\begin{multline}
j_\gamma(\epsilon_\gamma)=\int\rho_\gamma(\epsilon_\gamma)
\frac{dN_\pi}{dp_\perp dy}R_{pp}(E_p)\,dE_p\,dp_\perp\,dy\\
\text{photons}\;\text{cm}^3\;\text{s}^{-1}\;\text{MeV}^{-1}\;.
\end{multline}

The charged $\pi$'s decay to leptons, which can themselves be a source
of radiation from synchrotron and Compton processes.  We follow a
procedure for the $e^\pm$ completely analogous to that developed above
to find the steady state $e^\pm$ distribution:  
\begin{equation}
\dot{\rho}_e(E_e)=-2\kappa_e\,E_e\,\rho_e(E_e)-\kappa_e\,
E_e^2\,\frac{\partial\rho_e(E_e)}{\partial E_e}\;.
\end{equation}
This equation is easy to solve for the distribution $\rho_e(E_e)$,
which we then use to calculate the synchrotron and inverse Compton
spectra from the cascade $e^\pm$s.  Here, $\kappa_e\equiv(4/3)\sigma_Tc/(m_ec^2)^2(U_B+U_\gamma)$.

\section{Results and Discussion}
When the shock is located at $r_{sh}=40r_g$, we have $B\approx 435$
Gauss and $n_p\approx 1.6\times10^9 \;\text{cm}^{-3}$.  The
Rayleigh-Jeans tail cuts off at $\nu_{max}\approx10^{13}$ Hz, with a
temperature $\approx 6.3\times10^9$ K.  Thus,
$\gamma_{p,max}\approx1.2\times10^9$.  We find that for the
environment of Sgr A*, unlike those of typical AGNs, the $p$-$p$
collisions dominate over $p$-$\gamma$ by at least ten orders of
magnitude over the entire energy range.  This is because of the
relatively low density of ambient baryons, and the extreme dearth of
photons due to the low value of $\nu_{max}$.  Because of this, we
neglect the contribution from all $p$-$\gamma$ interactions.  We can
similarly neglect $p$-$e$ collisions for this system, the
cross-section of which is $\approx \frac{1}{137}\sigma_{pp}$.  Since
the population of relativistic $e^-$'s is much smaller than that of
ambient $p$'s, the contribution is insignificant.  The
synchrotron cooling channel dominates at the highest energy, and is
the other significant contributor to the spectrum.  From Sikora et
al. (1989), we find that any relativistic $n$'s with Lorentz factor
$\gamma_n\simlt10^8$ will escape the system without interacting, so we
consider as lost any $n$'s produced in the cascade.
The low photon density also means the region is extremely optically
thin to high-energy photons, so we consider the spectrum observed at
Earth to be that produced at the shock.
By comparison, $\gamma_{e,max}\approx 6.5\times 10^5$ for the $e^-$'s,
corresponding to an energy of $3.3\times10^5$ MeV.  For a roughly
equal injection rate of relativistic $e^-$'s and $p$'s, this energy
content is significantly below that of the cascade $e^-$'s, which
therefore dominate the leptonic contribution to the photon spectrum.

In all, five spectral components may be contributing 
to 2EGJ1746-2852: $p$ synchrotron, $p$ inverse Compton
scattering, $e^\pm$ synchrotron, $e^\pm$ inverse Compton scattering,
and $\pi^0$ decay.   For a reasonable efficiency (i.e.,
$\eta\simlt 10\%$), the $p$ synchrotron spectrum dominates over that
of the photons from $\pi^0$ decay as long as the $p$ injection index
$x \simlt 2.2$.  In Figure 1, we show these components for
the case when $r_{sh}=40r_g$, and $x=2.0$ with an efficiency of $1\%$.
The $p$ synchrotron seems to fit the data reasonably well, but clearly
misses the apparent low energy turnover in the EGRET data, and the
upper limits for the highest energy COMPTEL points.  This could be due
to the simplified geometry we have adopted in this paper, but this
is unlikely since the synchrotron spectrum depends primarily
on the particle physics.  It is also evident
in Figure 1 that Compton scattering is not important for this source,
and that the cascade $e^\pm$'s are relatively ineffective.  
The photons produced by $\pi$ decay are also
insignificant.  Placing the shock at $120r_g$ instead of $40r_g$
(Fig. 2) increases the emission area while decreasing $B$, and the $p$
synchrotron spectrum misses more data at the low end of the EGRET
spectrum.  The $e^\pm$ components still do not contribute.

With the shock at $40r_g$, the $\pi^0$ decay spectral component begins
to dominate over synchrotron when $x\simgt2.2$.  In Figure 3, we show
the same five components for the case $x=2.4$, which leads to $z=2.46$.  
Here, $\eta\approx 9\%$.
The shape of the $\pi^0$-induced $\gamma$-ray spectrum can be
understood as follows.  The center of the curve is set by the energy
($\epsilon_\gamma=67.5$ MeV) of the decay photons in the $\pi^0$ rest
frame, and the width is determined by Doppler broadening.  The slope
of the sides, and hence the index of the EGRET spectrum, is due to the
falloff in the number of decaying $\pi$'s at higher energy.  Each
$\pi$ decay produces a flat photon spectrum (see Eq. \ref{flat}) whose
width increases with $E_\pi$.  So the cumulative effect of all the
decays is greatest near $\epsilon_\gamma=67.5$ MeV,
where all $\pi$'s contribute.  The relative contribution to the
spectrum at lower or higher $\epsilon_\gamma$ then depends on the
overall $\pi$ distribution, which in turn is a function of both
$M_\pi$ and $\rho_p(E_p)$.  The flattened top,
and hence the low-energy turnover in the $\pi$ decay spectrum, is due
to the cutoff in $\pi$ production near the threshold.  The $\pi$ decay
spectrum cannot be translated laterally, so a simultaneous
match of both this turnover and the spectral slope is significant.

\begin{figure}[H] 
\centerline{\begin{turn}{-90}\epsfig{file=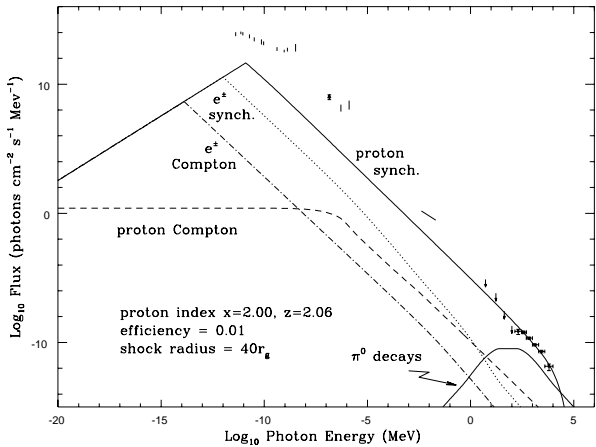,width=2.5in}\end{turn}}
\vspace{10pt}
\caption{Five spectral components (as labeled) resulting from the $p$-$p$ cascade.
The shock is here located at $40r_g$, and the injected proton spectral
index is $x=2.00$, yielding a steady state proton index $z=2.06$.
The inferred efficiency in this case is $\eta=1\%$.
The data points are from: (radio) Lo (1987), Zylka et al. (1993); (IR) 
Eckart et al. (1993), Stolovy (1996); (X-rays) Pavlinskii et
al. (1992);  ($\gamma$-rays) Strong 1996, Mattox (1996).}
\label{fig1}
\end{figure}

The required efficiency for producing the EGRET spectrum is different
depending on whether synchrotron or $\pi$ decay emissivity dominates.
There are three exit channels for the processed $p$ energy:
(i) $p$ synchrotron, (ii) $\pi^0\rightarrow\gamma\gamma$, and (iii)
$\pi^\pm\rightarrow \mu^\pm\nu_\mu$, with $\mu^\pm\rightarrow 
e^\pm\nu_e \nu_\mu $.  The latter two are coupled by a fixed ratio since
each of the three types of $\pi$'s receives an equal fraction of
the energy lost by the colliding $p$'s.  When $x\simlt2.2$,
(i) dominates and the conversion of $p$ power to
photons is very efficient.  When $x\simgt 2.2$, (ii) \& (iii) 
dominate, but the fraction of $p$ power
going into photons rather than particle by-products
is now smaller, so a higher $\eta$ is needed. 

\begin{figure}[H] 
\centerline{\begin{turn}{-90}\epsfig{file=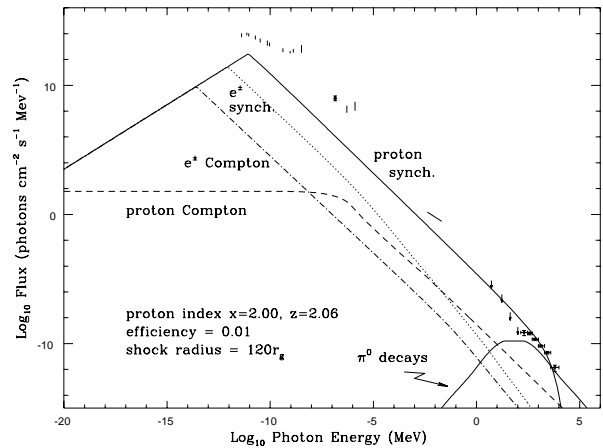,width=2.5in}\end{turn}}
\vspace{10pt}
\caption{Same as Fig. 1, except for a shock located at $120r_g$.}
\label{fig2}
\end{figure}

A population of relativistic
$p$'s energized within an accretion shock near a super-massive black
hole at the Galactic center may be contributing to the $\sim 30$
MeV--$10$ GeV emission from this region.  Depending on the value 
of the injected $p$ index (i.e., $x\simlt 2.2$ or $x\simgt 2.2$), this
contribution may come either from synchrotron or $\pi^0$
decay.  However, the synchrotron emissivity cannot account for
the turnover in the EGRET spectrum and some of the COMPTEL upper limits
at $\epsilon_\gamma\sim 100$ MeV, whereas the $\pi$-induced photon
distribution has a natural flattening there due to the threshold for
$\pi$ production in $p$-$p$ scatterings.  We have
not yet undertaken a detailed $\chi^2$ fitting to find the 
optimal parameters.  However, it may be an indication of robustness
in the model that a good fit was obtained without any fine-tuning. 
A more sophisticated fitting using the actual data will be 
undertaken in the future.
Our methods and results differ significantly from earlier treatments.  
Our use of $M_\pi$ and a careful 
treatment of the particle cascade give a good fit to 
the $\gamma$-ray spectrum with reasonable parameters, such 
as $z\approx 2.5$, $\eta\approx 0.09$, and a black hole mass $M_h\sim 10^6\;M_\odot$ 
(cf. \cite{mo94} who concluded that $z\approx 1.7$, and $M_h
\ll 10^6\;M_\odot$).

\begin{figure}[H] 
\centerline{\begin{turn}{-90}\epsfig{file=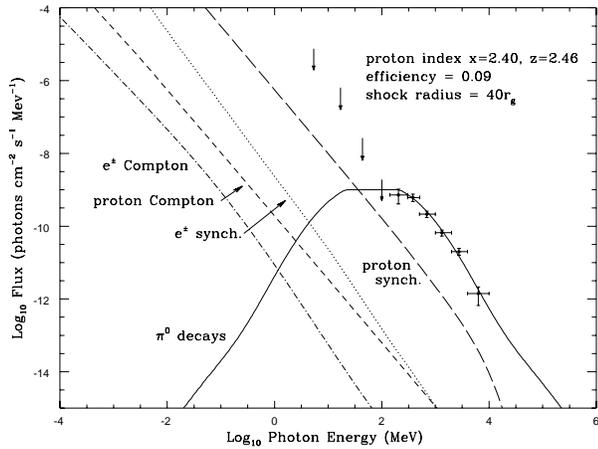,width=2.5in}\end{turn}}
\vspace{10pt}
\caption{Same as Fig. 1, except that the spectrum is here magnified to
highlight the EGRET and COMPTEL energy range.  In addition, the proton
injection index is $x=2.4
$, yielding a steady state proton index
$z=2.46$.  The inferred efficiency is here $\eta=9\%$.}
\label{fig3}
\end{figure}

\section{Acknowledgments}
This work was supported by an NSF Graduate Fellowship, and the NASA
grant NAGW-2518.  We acknowledge helpful discussions with
R. Jokipii and J. Mattox.
%
%
%

{}
\nopagebreak

\begin{thebibliography}{}

\bibitem[Abe et al.~1988]{abe88}\refindent Abe, F. et al. 1988, {\it PRL},
{\bf 61}, 1819.
\bibitem[Albajar et al.~1990]{alb90}\refindent Albajar, C. et al. 1990,
{\it Nuc. Phys. B}, {\bf 335}, 261.
\bibitem[Alner et al.~1986] {al86}\refindent Alner, G. et
al. 1986. {\it Zeits.Phys.}, {\bf 33C}, 1.
\bibitem[Alper et al.~1975]{alp75}\refindent Alper, B. et al. 1975,
{\it Nuc. Phys. B}, {\bf 100}, 237.
\bibitem[Babul et al.~1989] {ba89} \apj{Babul, A., Ostriker, J. \&
M\'{e}sz\'{a}ros, P. 1989}{347}{59} 
\bibitem[Barnett et al.~1996]{ba96}\refindent Barnett, R. et al. 1996,
{\it Phys.Rev.}, {\bf D54}, 1.
\bibitem[Begelman et al.~1990] {be90} \apj{Begelman, M., Rudak, B. \&
Sikora, M. 1990}{362}{38}
\bibitem[Churazov et al.~1994]{chu94}\apjsup{Churazov, E., et al. 1994}{92}{381}.
\bibitem[Cline 1988]{cli88} \refindent Cline, D. 1988, Proceedings of the RAND
Workshop on Antiproton Science and Technology, Ed. B.W. Augenstein,
et al. (New Jersey: World Scientific), 45.
\bibitem[Eckart et al.~1993]{ec93}\apjlett{Eckart, A. et al. 1993}{407}{L77}
\bibitem[Eckart \& Genzel 1997] {eg97} \mnras{Eckart, A. \& Genzel, R. 1997}{284}{576}
\bibitem[Fowler et al.~1984]{fo84}\refindent Fowler, G. et al. 1984,
{\it Phys.Lett.}, {\bf 145B}, 407.
\bibitem[Haller et al.~1996] {ha96} \apj{Haller, J. et al. 1996}{468}{955}
\bibitem[Jones \& Ellison 1991]{je91}\refindent Jones, F. \& Ellison, D. 1991, {\it
SSRv}, {\bf 58}, 259.
\bibitem[Lo 1987]{lo87}\refindent Lo, K.Y. 1987, in AIP Proc. 155,
ed. D.C. Backer (AIP: New York), {\bf 155}, 30.
\bibitem[Mahadevan et al.~1997]{mah97}\refindent Mahadevan, R., Narayan,
R. \& Krolik, J. 1997, astro-ph , 9704018.
\bibitem[Mastichiadis \& Ozernoy 1994]{mo94}\apj{Mastichiadis, A. \&
Ozernoy, L. 1994}{426}{599}
\bibitem[Mattox 1996]{mat96}\refindent Mattox, J.R. 1996, {\it GCNEWS
http://www.astro.umd.edu/~gcnews/gcnews/Vol.4/article.html}, {\bf 4}.
\bibitem[Mattox et al.~1996]{mae96}\apj{Mattox, J.R. et al. 1996}{461}{396}
\bibitem[Melia 1994]{mel94}\apj{Melia, F. 1994}{426}{577}
\bibitem[Menten et al.~1997] {men97} \apjlett{Menten, K., Reid, M.,
Eckart, A. \& Genzel, R. 1997}{475}{L111}
\bibitem[Merck et al. 1996]{mer96}\aasup{Merck, M. et al. 1996}{120}{465}.
\bibitem[Pavlinskii et al.~1992]{pav92}\sovlett{Pavlinskii, M., et al. 1992}{18}{291P}. 
\bibitem[Pohl 1997]{po97}\aa{Pohl, M. 1997}{317}{441}.
\bibitem[Predehl \& Tr\"umper 1994]{pred94}\aa{Predehl, P. \& 
Tr\"umper, J. 1994}{290}{L29}.
\bibitem[Ruffert \& Melia 1994]{ruf94}\aa{Ruffert, M. \& Melia, F. 1994}
{288}{L29}.
\bibitem[Rybicki \& Lightman 1979]{rl79}\refindent Rybicki, G. \&
Lightman, A. 1979, Radiative Processes in Astrophysics, (New York:
Wiley), 179 \& 207.
\bibitem[Sikora et al.~1989] {si89} \apjlett{Sikora, M. Begelman, M. \&
Rudak, B. 1989}{341}{L33}
\bibitem[Skinner et al. 1987]{skin87} \nature{Skinner, G.K., Willmore, A.P.,
Eyles, C.J., Bertram, D. \& Church, M.J. 1987}{330}{544}
\bibitem[Stolovy et al.~1996]{st96}\apjlett{Stolovy, S.R., Hayward, T.L. \& Herter, T.
1996}{470}{L45}
\bibitem[Strong 1996]{str96}\refindent Strong, A. W. 1996, private
communication to Pohl, M..
\bibitem[Watson et al. 1981]{wat81} \apj{Watson, M.G., Willingale, R., Grindlay, 
J.E., \& Hertz, P. 1981}{250}{142}
\bibitem[Zylka et al.~1993]{zyl93}\aa{Zylka, R., Mezger, P.G. \& Lesch, H. 1993}
{261}{119}.
\bigskip\bigskip
\end{thebibliography}
\end{document}